\documentclass[11pt,a4paper]{article}
\pdfoutput=1
\usepackage{jheppub}

\usepackage{multirow, graphicx,amssymb,url,mathrsfs,amsmath}
\usepackage{wrapfig,boxedminipage,setspace,subfigure,epsfig}
\usepackage{amsxtra,amstext,latexsym,dsfont,amsfonts}
\usepackage{color,eucal}
\usepackage[dvipsnames]{xcolor}
\usepackage{float}
\usepackage{slashed,comment}
\usepackage{kotex}
\usepackage{tensor}

\usepackage{indentfirst}

\newcommand{\dd}{\mathrm{d}}

\title{Holographic study of $T\bar{T}$ like deformed HV QFTs: \\ 
holographic entanglement entropy}

\author[a,b,c,d]{Hyun-Sik Jeong,}
\author[a]{Wen-Bin Pan,}
\author[a,b]{Ya-Wen Sun,}
\author[a]{and Yuan-Tai Wang}

\emailAdd{sicobysico@gmail.com}
\emailAdd{panwenbin18@mails.ucas.ac.cn}
\emailAdd{yawen.sun@ucas.ac.cn}
\emailAdd{wangyuantai19@mails.ucas.ac.cn}
 
\affiliation[a]{School of physics $\&$ CAS Center for Excellence in Topological Quantum Computation, University of Chinese Academy of Sciences, Zhongguancun east road 80, Beijing 100049, China}
\affiliation[b]{Kavli Institute for Theoretical Sciences, University of Chinese Academy of Sciences, \\ Zhongguancun east road 80, Beijing 100049, China}
\affiliation[c]{Instituto de Física Teórica UAM/CSIC, Calle Nicolás Cabrera 13-15, 28049 Madrid, Spain}
\affiliation[d]{Departamento de Física Teórica, Universidad Autónoma de Madrid, Campus de Cantoblanco, 28049 Madrid, Spain}

\abstract{
We study the $(d+2)$-dimensional Hyperscaling Violating (HV) geometries in the presence of both a finite temperature $T$ and a UV cutoff $r_c$. This gravitational system is conjectured to be dual to $T\bar{T}$ like deformed HV QFTs. We consider the representative quantum entanglement quantity in holography, i.e. the entanglement entropy $S(A)$, and perform a complete analysis in all possible parameter ranges of the hyperscaling violation exponent $\theta$ and the critical dynamical exponent $z$ to study the effect of the temperature and the cutoff. 
We find that the temperature has a universal effect independent of the parameters: it enhances $S(A)$ in the small cutoff limit, while it is irrelevant in the large cutoff limit. 
For the cutoff effect, we find that the cutoff monotonically suppresses $S(A)$ where its behavior depends on the parameter range. 
As an application of the finite temperature analysis, we study the first law of entanglement entropy, $S_{T}-S_{T=0}\sim\ell^{\lambda}$, in the small subsystem size $\ell$ limit. We find that $\lambda$ interpolates between $\lambda=1+z$ in the small cutoff and $\lambda=3$ in the large cutoff, independent of the parameter range.
We also provide the analytic holographic result at $z=d-\theta$ and discuss its possibility of comparison with the field theoretic result.

}

\begin{document}
\maketitle

\section{Introduction}
\label{sec:introduction}

The AdS/CFT correspondence (or gauge/gravity duality)~\cite{Maldacena:1997re,Gubser:1998bc,Witten:1998qj} is a conjectured duality between the AdS gravity theories and the conformal field theories (CFTs). 
It provides a new insight to better understand strongly coupled quantum field theories in terms of geometric quantities.
A representative example is the Ryu-Takayanagi formula~\cite{Ryu:2006bv,Ryu:2006ef}: the entanglement entropy of the boundary CFTs can be related to some geometric quantity in the AdS bulk spacetime. In other words, the Ryu-Takayanagi formula could indicate that spacetime may emerge from the entanglement properties of the boundary dual field theories. (see also \cite{Nishioka:2009un,VanRaamsdonk:2010pw,Nozaki:2012zj,Lin:2014hva,Hayden:2016cfa}.)

It may be instructive to note that the conformal symmetry often plays an important role in explicitly obtaining the AdS/CFT correspondence, e.g., in the computations of the partition function and correlation functions.
One of the recent developments for the AdS/CFT correspondence is to generalize such a correspondence by  deforming the conformal symmetry. In this paper we focus on two interesting approaches in this direction: in particular, for the gravity side by considering a) a finite radial cutoff; b) a larger class of geometries (i.e., beyond AdS), especially the hyperscaling violating geometry~\cite{Charmousis:2010zz}.

\paragraph{AdS geometry with a cutoff (corresponding to $T\bar{T}$ deformed CFTs):}
with \textit{a finite radial cutoff} in the AdS geometries, it was proposed \cite{McGough:2016lol} that the $T\bar{T}$ deformed CFTs could be a dual field theory of this gravity system, whose action follows\footnote{For instance, for the two-dimensional CFTs, one can find $T:=T_{ww},\, \bar{T}:=T_{\bar{w}\bar{w}}$ in which $w=x+i\tau$ and $\bar{w}=x-i\tau$ are complex coordinates ($x,\,\tau$).}
\begin{align}\label{OPXAC}
\begin{split}
\frac{\partial S(\xi)}{\partial \xi} = \int \dd^{d+1} x \sqrt{g} \, X(x) \,,
\end{split}
\end{align}
where $\xi$ is the deformation parameter and $X$ is the $T\bar{T}$ deformation operator~\cite{Taylor:2018xcy,Hartman:2018tkw,McGough:2016lol,Kraus:2018xrn,Cardy:2018sdv,Bonelli:2018kik,Gross:2019ach,Gross:2019uxi} 
\begin{equation}\label{OPX}
    X = T^{\mu\nu}T_{\mu\nu} - \frac{1}{d}\left(T_\mu^\mu\right)^2.
\end{equation}
The CFT parameter $\xi$ is related to the bulk UV radial cutoff $r_c$ in the dual gravity system
\begin{align}
\begin{split}
\xi = \frac{4\pi G_{N}}{d+1}\,r_c^{d+1} \,.
\end{split}
\end{align}
Here $G_{N}$ is the gravitational constant and $T_{\mu\nu}$ the stress tensor in the $T\bar{T}$ deformed CFT.\footnote{One interesting aspect of the operator $X$ Eq.\eqref{OPX} is the factorization property~\cite{Zamolodchikov:2004ce,Taylor:2018xcy,Hartman:2018tkw,McGough:2016lol,Grieninger:2019zts}, which may be related to the calculation of energy spectrum~\cite{Zamolodchikov:2004ce,Smirnov:2016lqw,Cavaglia:2016oda}.}

Using the $T\bar{T}$ deformed CFTs Eq.\eqref{OPX}, various aspects of it have been extensively investigated both on the field theory side and the gravity side. For instance, the correlation functions~\cite{Kraus:2018xrn,Aharony:2018vux,Cardy:2019qao,He:2019vzf,He:2019ahx,He:2020udl,He:2020cxp}, the entanglement entropy~\cite{Donnelly:2018bef,Chen:2018eqk,Park:2018snf,Banerjee:2019ewu,Murdia:2019fax,Jeong:2019ylz,Grieninger:2019zts,Donnelly:2019pie},  mutual information~\cite{Asrat:2020uib}, and the entanglement wedge cross section~\cite{Asrat:2020uib}.

\paragraph{Hyperscaling Violating geometry (corresponding to HV QFTs):} another attempt to study the generalized AdS/CFT correspondence is to consider \textit{a large class of special geometries}~\cite{Charmousis:2010zz}, i.e. the Hyperscaling Violating geometry (HV geometry), as 
\begin{align}\label{HVMETRIC}
\begin{split}
\dd s^2 =&\,\, r^{-\frac{2\,d_e}{d}} \left( -r^{-2(z-1)} \dd t^2 + \dd r^2 + \dd x_{i}^2 \right) \,, \\ 
 d_{e} :=&\,\, d-\theta \,,
\end{split}
\end{align}
where $i=1,2,\dots,d$. This metric Eq.\eqref{HVMETRIC} contains two geometric parameters: the dynamical critical exponent $z$ and the hyperscaling violation exponent $\theta$. Note that Eq.\eqref{HVMETRIC} becomes the AdS metric when $(z,d_e)=(1,d)$ or equivalently $(z,\theta)=(1,0)$.\footnote{One can easily find that the metric Eq.\eqref{HVMETRIC} is scale invariant up to an overall scaling of the metric, i.e., $\dd s \rightarrow \xi^{\theta/d} \dd s$ under the scaling $(t,\, x_i,\, r)\rightarrow(\xi^z \, t,\, \xi \, x_i,\, \xi \, r)$.}

Note that the study of this larger class of geometries Eq.\eqref{HVMETRIC} is motivated by the investigation of condensed matter systems using the gravitational toy model.
For instance, in holography, it was proposed \cite{Blake:2016sud,Blake:2016jnn,Blake:2017qgd,Ahn:2017kvc} that the energy diffusivity ($D_e$) may be related to the quantum chaos properties (butterfly velocity $v_B$, Lyapunov time $\tau_L$) with a universal relation, i.e., the ratio between them, ${D_e}/\left({v_{B}^2\,\tau_{L}}\right){=z/2(z-1)}$, is only a function of the dynamical critical exponent $z$ at low temperature, independent of other exponents or UV data such as the momentum relaxation strength.

Note also that such a universal relation was first proposed in holographic models, and it has also been observed in condensed matter theories~\cite{Zhang:2016ofh,Aleiner:2016aa,Swingle:2016aa,Patel:2016aa,Bohrdt:2016vhv,Werman:2017abn} as well as in the Sachdev-Ye-Kitaev (SYK) models~\cite{Davison:2016ngz,Gu:2016oyy,Jian:2017unn}. See also \cite{Blake:2018leo,Jeong:2021zhz,Jeong:2022luo} for its relation with the ill-defined Green's function, pole-skipping phenomena.

\paragraph{Hyperscaling Violating geometry with a cutoff (corresponding to $T\bar{T}$ like deformed HV QFTs):} as motivated by the proposal in \cite{McGough:2016lol} of AdS geometry with a UV cutoff, one may also study the role of a finite UV cutoff on the HV geometry Eq.\eqref{HVMETRIC}, i,e., \textit{the larger classes of geometries with a finite radial cutoff}.

 AdS geometry with a finite UV cutoff corresponds to $T\bar{T}$ deformed CFTs on the boundary. One may then ask which quantum field theories (QFTs) are dual to the HV geometry with a cutoff. In other words, what is the deformation operator $X$ in Eq.\eqref{OPXAC} for generic $(z, d_e)$?

In \cite{Alishahiha:2019lng}, it was proposed that such a dual field theory may be $T\bar{T}$ like deformed HV QFTs. In particular, using the renormalized Brown-York stress tensor for the HV geometries in addition to the proper counter terms, it was shown in \cite{Alishahiha:2019lng} that the corresponding deformation operator may be given by 
\begin{align}\label{OPXHV}
\begin{split}
X \,=\, & z \left(T_{t}^{t}\right)^2 \,+\, \frac{d_e}{d} T_{j}^{i}\,T_{i}^{j} \,-\, \frac{1}{d_e} \left( z T_{t}^{t} + \frac{d_e}{d} T_{i}^{i}  \right)^2  \\
& \,-\, \frac{z(z-1)}{2 d_e^2}\left(z T^t_t+\frac{d_e}{d}T^i_i\right)J^t A_t + \frac{z(d_e+z)(d_e+z-1)}{2d_e^2} T^t_t J^t A_t \,,
\end{split}
\end{align}
which reduces to the pure AdS case Eq.\eqref{OPX} when $(z,\theta)=(1,0)$.
{Note that Eq.\eqref{OPXHV} is defined at finite cutoff $r_c$, for instance, the current $J^t\sim\sqrt{z-1}(\sqrt{f(r_c)}-1)$: also notice that one can find the contribution of the current (or the gauge field $A_\mu$) only at non-zero temperature, i.e, when $f(r)\neq1$ in Eq.\eqref{HVMETRIC2}.}
See \cite{Alishahiha:2019lng} for details.\footnote{It is speculated that Eq.\eqref{OPXHV} also has the factorization property as in Eq.\eqref{OPX}. See \cite{Khoeini-Moghaddam:2020ymm} and the references therein.}

\paragraph{Motivation of this paper:} In this paper, we aim to explore the properties of the $T\bar{T}$ like deformed HV QFTs in holography, i.e., HV geometries with a finite cutoff.  In particular, in order to investigate the properties of such a geometry, we focus on the most representative holographic entanglement measure: \textit{entanglement entropy}.\footnote{For the holographic study of other measures for the mixed states, such as mutual information and the entanglement wedge cross section, in the same setup, see \cite{wipYW3}.} 

Note that our work could be an extended analysis of previous studies of the $T\bar{T}$ deformed CFTs ($z=1,d_e=d$) in holography~\cite{Donnelly:2018bef,Chen:2018eqk,Park:2018snf,Banerjee:2019ewu,Murdia:2019fax,Jeong:2019ylz,Grieninger:2019zts,Donnelly:2019pie}.
Moreover, as did in the case of the universal relation of the diffusivity, our holographic work may also be compared with other theories in future.

For a systematic and complete study of the entanglement entropy in the whole range of HV parameters ($z, d_e$), we also consider the finite temperature contribution by adding the emblackening factor~\cite{Dong:2012se}, $f(r)$, on the metric Eq.\eqref{HVMETRIC} as
\begin{align}\label{HVMETRIC2}
\begin{split}
\dd s^2 =&\, r^{-\frac{2\,d_e}{d}} \left( -r^{-2(z-1)} \, f(r)\, \dd t^2 + \frac{\dd r^2}{f(r)} + \dd x_{i}^2 \right) \,, \\ 
f(r)=&\,1-\left(\frac{r}{r_{h}}\right)^{d_e+z} \,,
\end{split}
\end{align}
where $d_e$ is given in Eq.\eqref{HVMETRIC}, and $r_h$ is the horizon radius. The Hawking temperature ($T$) for Eq.\eqref{HVMETRIC2} is given by 
\begin{align}\label{HAWKT}
\begin{split}
T:=\frac{r_{h}^{1-z}\,|f'(r_h)|}{4\pi}=\frac{1}{4 \pi} \frac{|d_e+z|}{r_{h}^{z}} \,.
\end{split}
\end{align}
Note that a finite $T$ (or $f(r)\neq1$) is required to study the effect of the dynamical critical exponent $z$ on the entanglement properties, for instance, at fixed time ($\dd t =0$) one may not find a $z$-dependence on the entanglement entropy when $f(r)=1$.

Using the null energy condition as well as the positive specific heat condition, one can also find the physically relevant parameter range for ($z, d_e$)~\cite{Dong:2012se} as
\begin{align}\label{RANGECON}
\begin{split}
d_e=0:& \quad z\leq0 \text{\quad or\quad} z\geq1 \,, \\ 
0<d_e<1:& \quad z \geq 2 - \frac{d_{e}}{d} \,, \\
d_e\geq1:& \quad 1 \leq d_e < d \text{\,\,\,and\,\,\,} z \geq 2-\frac{d_e}{d}    \text{\quad or \,\,\,\,} d_e \geq d \text{\,\,\,and\,\,\,} z \geq 1 \,,
\end{split} 
\end{align}
where $d\geq1$. Note that $d_e<0$ is forbidden in that the behavior of entanglement entropy is not consistent with the one in QFTs~\cite{Dong:2012se}.\footnote{It also turned out $d_e<0$ may be related to the instabilities in the gravity theories without the well defined decoupling limits in the string theory realization~\cite{Dong:2012se}.}

Without a cutoff, an interesting behavior of the entanglement entropy has been found for the second parameter range above, $0<d_e<1$, in Eq.\eqref{RANGECON}: the violations of the area law~\cite{Dong:2012se}.
Moreover, at the limit $d_e=0$, i.e. the first parameter range in Eq.\eqref{RANGECON}, the volume law behavior for the entanglement entropy may appear, which is related to an extensive contribution to the entanglement entropy. The opposite limit $d_e=1$ (the lower bound of the third parameter range in Eq.\eqref{RANGECON}) is also of interest in that the holographic theories may give the properties of Fermi surfaces~\cite{Ogawa:2011bz,Huijse:2011ef}.\footnote{As the thermal entropy is proportional to temperature as $T^{d_{e}/z}$, $d_e$ may play the role of an effective space dimension for the dual field theory~\cite{Dong:2012se}. In this sense, $d_e=0$ yields a system living in the (0+1) dimension, while $d_e=1$ may correspond to the system living in the ($1+1$) dimension which may be consistent with the interpretation of $d_e=1$ system as possessing a fermi surface.}
In the main context, considering all the cases in Eq.\eqref{RANGECON}, we extend the analysis in \cite{Dong:2012se} in the presence of a finite UV cutoff. For instance, we explore the effect of the cutoff on the violation of the area law.

It is also instructive to note that the entanglement entropy of the HV geometries at finite cutoff has been investigated in previous literature~\cite{Khoeini-Moghaddam:2020ymm,Paul:2020gou}.\footnote{See also \cite{Alishahiha:2019lng} for the study of energy spectrum and holographic complexity. The analysis of the Page curve in the HV geometries can be seen in \cite{Omidi:2021opl}.} 
However, to our knowledge, the effect of a finite temperature as well as considering the full parameter range Eq.\eqref{RANGECON} has not been investigated yet. For instance, only the third parameter range in Eq.\eqref{RANGECON}, $d_e\geq1$, was investigated at zero temperature in \cite{Khoeini-Moghaddam:2020ymm,Paul:2020gou}.

Thus, in this paper, considering a finite temperature, we analyze the effect of a finite cutoff on the entanglement properties with the full HV parameters ($z, d_e$) within the complete parameter range Eq.\eqref{RANGECON}. Moreover, as an application of the study at finite temperature, we also discuss the properties of the thermal excitation of the entanglement entropy: the first law of entanglement entropy (FLEE)~\cite{Bhattacharya:2012mi}.\footnote{For recent developments of FLEE with spontaneously broken U(1) or translational symmetry, see \cite{Jeong:2022zea}  with the references therein.}

This paper is organized as follows. 
In section \ref{sec3}, we briefly review the Ryu-Takayanagi formula to compute entanglement entropy for the case of the strip subsystem. Then we investigate holographic entanglement entropy of the HV geometries at finite temperature/cutoff. 
In section \ref{sec4}, we discuss two applications of the holographic entanglement entropy at finite temperature, including FLEE.
Section \ref{sec5} is devoted to conclusions.

\section{Holographic entanglement entropy}
\label{sec3}

In this section, we study the holographic entanglement entropy, $S(A)$, in Hyperscaling Violating (HV) geometries at finite temperature and finite radial cutoff.

\subsection{Preliminary}\label{sec3.1}

We first review $S(A)$ in HV geometries at zero temperature~\cite{Khoeini-Moghaddam:2020ymm} and extend the analysis to the case at finite temperature in which, in general, $S(A)$ in Eq.\eqref{HEEFOR} cannot be evaluated analytically so that one needs to resort to  numerical analysis.

\paragraph{Holographic entanglement entropy (HEE):} The entanglement entropy is one of the well studied entanglement measures in quantum information theory. It is defined as the von Neumann entropy of the reduced density matrix for the subsystem under consideration and can be used to characterize the entanglement of the bipartite \textit{pure} state.

In holography, the gravity dual of the entanglement entropy is the Ryu-Takayanagi formula~\cite{Ryu:2006bv,Ryu:2006ef}, which corresponds to the computation of the minimal area of the bulk surface (or Ryu-Takayanagi surface) anchored on the boundary region of interest.
For instance, the holographic entanglement entropy of a subsystem $A$, $S(A)$, is given as
\begin{align}\label{RT}
S(A)=\frac{\text{Area}(\Gamma_A)}{4G_N}\,,
\end{align}
where $\Gamma_A$ is the minimal bulk surface at a fixed $t$ and $G_N$ is the Newton constant.

\begin{figure}[]
\centering
     \includegraphics[width=7.5cm]{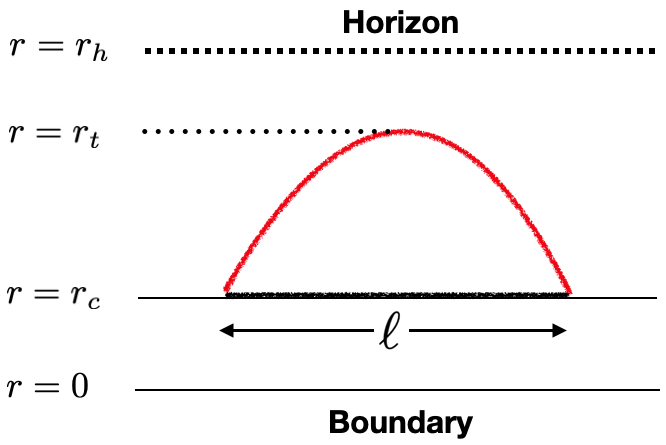}
 \caption{A strip entangling region with the Ryu-Takayanagi surface (red). The strip has the length $\ell$ in the $x_1$-direction. The length in other directions is denoted as $L:=\int \dd x_{i=2,\dots,d}$. $r_c$ is the cutoff, $r_t$ the largest $r$ value of the minimal surface in the bulk, and $r_h$ the black hole horizon.}\label{RTCT}
\end{figure}

In this paper, we consider the strip entangling surface. See Fig. \ref{RTCT}: the Ryu-Takayanagi surface $\Gamma_A$ is expressed as the red surface with the length of the strip $\ell$.

With the metric of our interest, Eq.\eqref{HVMETRIC2}, HEE Eq.\eqref{RT} could be obtained as
\begin{align}\label{HEEFOR}
\begin{split}
    S(A)=\,\,\frac{L^{d-1}}{2G_N}\int^{r_t}_{r_c} \frac{r^{-d_e}}{\sqrt{\left(1-\left(\frac{r}{r_h}\right)^{d_{e}+z}\right)\left(1-\left(\frac{r}{r_t}\right)^{2d_e}\right)}} \,\dd r \,, \quad 
     L^{d-1}:=\,\, \int \dd x_2\dots x_d \,,
\end{split}
\end{align}
together with the length of the strip $\ell$
\begin{align}\label{lengthfor}
    \ell = 2 \int_{r_{c}}^{r_{t}} \frac{ \left(\frac{r}{r_t}\right)^{d_{e}}}{\sqrt{\left(1-\left(\frac{r}{r_h}\right)^{d_{e}+z}\right)\left(1-\left(\frac{r}{r_t}\right)^{2 d_{e}}\right)}} \,\dd r \,,
\end{align} where $r_c$ is the cutoff, $r_t$ is the largest value of $r$ of the minimal surface in the bulk, and $r_h$ is the black hole horizon.

\paragraph{Zero temperature:}
In order to study the case at zero temperature, one may consider $r_h\rightarrow\infty$ as could be seen from Eq.\eqref{HAWKT}. Then the holographic entanglement entropy Eq.\eqref{HEEFOR} for this case could be computed analytically as 
\begin{align}\label{Sfor}
\begin{split}
S(A)=
\begin{cases}
(d_e \ne 1):& \displaystyle{\frac{L^{d-1}}{2 G_{N} \left(d_{e}-1\right)}\Bigg[-\frac{\Gamma_1}{r_{t}^{d_{e}-1}}+\frac{1}{r_{c}^{d_{e}-1}} \, { }_{2} F_{1}\left[\frac{1}{2}, \frac{1-d_{e}}{2 d_{e}}, \frac{d_{e}+1}{2 d_{e}},\left(\frac{r_{c}}{r_{t}}\right)^{2 d_{e}}\right]\Bigg] \,,} \\
(d_e = 1):& \displaystyle{\frac{L^{d-1}}{2 G_{N} }\log\frac{r_t+\sqrt{r_t^2-r_c^2}}{r_c}  \,, }
\end{cases}
\end{split}
\end{align}
where ${ }_{2} F_{1}$ is the hypergeometric function and we have defined 
\begin{align}\label{GAMMA1F}
    \Gamma_1 := \frac{\sqrt{\pi} \,  \Gamma\left(\frac{d_{e}+1}{2 d_{e}}\right)}{\Gamma\left(\frac{1}{2 d_{e}}\right)} \,.
\end{align}
Also, the length of the strip Eq.\eqref{lengthfor} is determined as 
\begin{align}\label{lfor}
\begin{split}
\frac{\ell}{2}=
\begin{cases}
    (d_e \ne 1):& \displaystyle{ \Gamma_1 r_{t}-\frac{r_{c}^{d_{e}+1}}{\left(d_{e}+1\right) r_{t}^{d_{e}}} \,}  \,\, \displaystyle{{ }_{2} F_{1}\left[\frac{1}{2}, \frac{d_{e}+1}{2 d_{e}}, \frac{3 d_{e}+1}{2 d_{e}},\frac{r_{c}^{2 d_{e}}}{r_{t}^{2 d_{e}}} \right] \,,} \\
     (d_e = 1):& \displaystyle{ \sqrt{r_t^2-r_c^2} \,, }
\end{cases}
\end{split}
\end{align}

Note that dialing $r_t$ at fixed ($r_c, \, d_e$), one could also express $S(A)$ in terms of the length scale $\ell$. Such an expression could be useful in studying some properties of the entanglement entropy, for instance, the first law of entanglement entropy as we will do.
Note also that Eq.\eqref{Sfor}-Eq.\eqref{lfor} are the reproduction of the zero temperature results in the HV geometries \cite{Khoeini-Moghaddam:2020ymm}.

{
We consider the case $d_e>0$ in the main text because when $d_e=0$ the metric Eq.\eqref{HVMETRIC2} at fixed time is flat  so that the Ryu-Takayanagi surface lies completely on the boundary $r=r_c$ and coincides with the strip.
Since the strip has an area of $L^{d-1} \ell$, the HEE is $S_{d_e=0}(A)=\frac{L^{d-1} \ell}{4 G_N}$. Thus it receives no correction from the cutoff or the temperature in the case of $d_e=0$.
}

\paragraph{Finite temperature:}
Now we extend the analysis in  \cite{Khoeini-Moghaddam:2020ymm} to the finite temperature case in which the corresponding metric becomes Eq.\eqref{HVMETRIC2} with a finite $r_h$.

There are two major features distinct from the zero temperature case. First, unlike the zero temperature case, $S(A)$ Eq.\eqref{HEEFOR} and $\ell$ Eq.\eqref{lengthfor} cannot be evaluated analytically in general. Second, at finite temperature, one could also find the effect of the dynamical critical exponent $z$ on $S(A)$. Recall that such $z$ contribution originates from the $g_{rr}$ component of the metric Eq.\eqref{HVMETRIC2} when $f(r)\neq1$.

For the case of a finite temperature, we have two ``HV'' parameters and three ``Bulk'' parameters as
\begin{align}\label{}
\text{HV parameters}:  (d_e, z) \,, \qquad  \text{Bulk parameters}:  (r_c, r_t, r_h) \,.
\end{align}
Strictly speaking, we also have additional parameters ($d, L, G_N$) for the study of $S(A)$, which would be irrelevant for our discussion in that we can rescale $S(A)$ as $S(A)2G_N/L^{d-1}$. Thus, to simplify our formulas and avoid clutter, we set $L=1$ and $G_N=1/2$ hereafter.

At given HV parameters, we focus on investigating the cutoff ($r_c$) and temperature dependence ($r_h$) of $S(A)$. For this purpose, we define the dimensionless parameters as 
\begin{align}\label{DMLS}
    \tilde{S} := \frac{S}{r_t^{1-d_e}} \,, \qquad \tilde{r}_{c} := \frac{r_c}{r_t} \,, \qquad
    \tilde{r}_{h} := \frac{r_h}{r_t} \,,
\end{align}
where $\tilde{r}_c < 1 < \tilde{r}_h$: recall that $r_c<r_t<r_h$, see Fig. \ref{RTCT}.
Furthermore, using the length of the strip Eq.\eqref{lengthfor} together with $\tilde{\ell} := {\ell}/{r_t}$, we can also  express Eq.\eqref{DMLS} at fixed subsystem size $\ell$ as
\begin{align}\label{}
    \bar{S} := \frac{\tilde{S}}{\tilde{\ell}^{1-d_e}} = \frac{S}{\ell^{1-d_e}} \,, \qquad \bar{r}_{c} := \frac{\tilde{r}_c}{\tilde{\ell}} = \frac{r_c}{\ell} \,, \qquad \bar{r}_{h} := \frac{\tilde{r}_{h}}{\tilde{\ell}} = \frac{r_h}{\ell} \,.
\end{align}

In what follows, for given HV parameters ($d_e, z$), we study the cutoff ($\bar{r}_c$) and the temperature ($\bar{r}_h$) dependence of the entanglement entropy $\bar{S}$. We also discuss the subsystem size ($\ell$) dependence in the section for the first law of entanglement entropy.

In the context, we mainly provide the analytic results of $S(A)$ in order not only to better understand the cutoff/temperature correction, but also to support our numerical results. Note that we will perform a complete analysis considering all possible parameter ranges Eq.\eqref{RANGECON} for HV parameters.

In particular, for the analytic treatment, we focus on the \textit{small temperature} correction ($\tilde{r}_h \gg 1$) within two cases:
\begin{itemize}
    \item {\textit{Small cutoff} case ($\tilde{r}_c \ll 1$),\, i.e.,\,\, $\tilde{r}_c \ll 1 \ll \tilde{r}_h$:\,\, Section \ref{sec3.2}.}
    \item {\textit{Large cutoff} case ($\tilde{r}_c \sim 1$),\,\, i.e.,\,\,\,  $\tilde{r}_c \,\sim 1 \ll \tilde{r}_h$:\,\, Section \ref{sec3.3}.}
\end{itemize}
In addition, we also provide two applications of the finite temperature analysis: i) the first law of entanglement entropy in Section \ref{sec3.4}; ii) the special case ($d_e=z$) in which the analytic analysis is available for any value of $\tilde{r}_c$ and $\tilde{r}_h$ in Section \ref{sec3.5}.

%
\subsection{Analytic results at low temperature} 

\subsubsection{Small cutoff analysis}\label{sec3.2}

Let us first start from the small cutoff analysis. In order to do this, one could expand the integrand in Eq.\eqref{HEEFOR} in the ($\tilde{r}_c \ll 1 \ll \tilde{r}_h$) expansion and integrate it order-by-order.
One could also check that our small cutoff condition ($\tilde{r}_c \ll 1 \ll \tilde{r}_h$) is equivalent to  
\begin{align}\label{SMCREVI}
   r_c \ll r_t \ll r_h    \quad\rightarrow\quad  r_c \ll \ell \ll r_h  \quad\rightarrow\quad  r_c/\ell \ll 1 \ll r_h/\ell  \,,
\end{align}
for instance, see Eq.\eqref{rte1} to find $r_t\sim\ell$.

As demonstrated in the introduction, $S(A)$ could be classified depending on the parameter regime Eq.\eqref{RANGECON} and we find  
\begin{align}\label{Sle1de}
\frac{S(A)}{\ell^{1-d_e}}=
\small{
\begin{cases}
\,{(0 < d_e < 1):}& \displaystyle{ \frac{2^{d_e-1}\Gamma_1^{d_e}}{1-d_e} \,-\, \frac{1}{1-d_e} \left(\frac{r_c}{\ell}\right)^{1-d_e} \,+\, \frac{d_e \Gamma_2}{2^{2+z}(1+z)(1+d_e+z)\Gamma_1^{1+z}} \frac{1}{\left( \frac{r_h}{\ell} \right)^{d_e+z}} \,, }    \\
 \qquad (d_e=1):& \displaystyle{ -\log{\frac{r_c}{\ell}} \,+\, \frac{d_e \Gamma_2}{2^{2+z}(1+z)(1+d_e+z)\Gamma_1^{1+z}} \frac{1}{\left( \frac{r_h}{\ell} \right)^{d_e+z}}\Big|_{d_e=1}  \,, }  \\
\qquad(d_e > 1):& \displaystyle{ \frac{1}{d_e-1} \frac{1}{\left(\frac{r_c}{\ell}\right)^{d_e-1}} \,+\, \frac{2^{d_e-1}\Gamma_1^{d_e}}{1-d_e} \,+\, \frac{d_e \Gamma_2}{2^{2+z}(1+z)(1+d_e+z)\Gamma_1^{1+z}} \frac{1}{\left( \frac{r_h}{\ell} \right)^{d_e+z}} \,, }
\end{cases}
}
\end{align}
where $\dots$ denotes higher order corrections, $\Gamma_1$ is Eq.\eqref{GAMMA1F} and $\Gamma_2$ is defined as
\begin{align}\label{GAMMA2F}
    \Gamma_2=\frac{\sqrt{\pi} \, \Gamma\left(1+\frac{1}{2 d_{e}}+\frac{z}{2d_e}\right)}{\Gamma\left(\frac{1}{2}+\frac{1}{2 d_{e}}+\frac{z}{2d_e}\right)} \,.
\end{align}
Note that the power ($d_e+z$) in Eq.\eqref{Sle1de} is always positive within the parameter range Eq.\eqref{RANGECON} so that the series expansion for the temperature is reliable.

Note that, in Eq.\eqref{Sle1de}, we keep terms up to the leading order of the cutoff, $r_c/\ell$, and the temperature corrections, $(r_h/\ell)^{-1}$, for instance $\dots$ may include terms like $(r_c/\ell)^{1+d_e}$.
Also, in order to express $S(A)$ at fixed subsystem size $\ell$ in Eq.\eqref{Sle1de}, we used
\begin{align}\label{rte1}
\begin{split}
    r_t \,=\, & \frac{\ell}{2\Gamma_1}  \Bigg[1 \,+\, \frac{2^{1+d_e}\Gamma_1^{d_e}}{1+d_e} \left(\frac{r_c}{\ell}\right)^{1+d_e} \,-\, \frac{\Gamma_2}{(1+d_e+z)(2\Gamma_1)^{1+d_e+z}}\left(\frac{\ell}{r_h}\right)^{d_e+z} \,+\, \dots \Bigg] \,,
\end{split}
\end{align}
which is obtained by expanding the integrand of $\ell$ Eq.\eqref{lengthfor}.

Note also that our results Eq.\eqref{Sle1de} not only extend the zero temperature result \cite{Khoeini-Moghaddam:2020ymm} to the finite temperature case, but also fills the gap in the literature: the first line in Eq.\eqref{Sle1de}, $(0<d_e<1)$ case, was not investigated even at zero temperature~\cite{Khoeini-Moghaddam:2020ymm}.\footnote{Our results Eq.\eqref{Sle1de} are also consistent with \cite{Dong:2012se} when we take the limit $r_c\rightarrow0$ (zero cutoff) and $r_h\rightarrow\infty$ (zero temperature). See also the discussion for the area/volume law therein, which are related to the first terms in each case of Eq.\eqref{Sle1de}.}

Next, let us first summarize the temperature/cutoff correction for $S(A)$ that we find
\begin{itemize}
    \item{The temperature \textit{enhances} $S(A)$;}
    \item{The cutoff \textit{suppresses} $S(A)$;}
\end{itemize}
which will be explained further below.

We find that the temperature correction is universal for all parameter range
\begin{align}\label{UTCF}
\frac{d_e \Gamma_2}{2^{2+z}(1+z)(1+d_e+z)\Gamma_1^{1+z}} \frac{1}{\left( \frac{r_h}{\ell} \right)^{d_e+z}} \,,
\end{align}
where this also implies that the temperature \textit{enhances} the entanglement entropy $S(A)$: we checked that the prefactor is always positive for all parameter range.
Note that a temperature correction is $1/(r_h^{d_e+z}) \sim T^{\frac{d_e+z}{z}}$ with $\frac{d_e+z}{z}>0$ given in the parameter range Eq.\eqref{RANGECON}. It may also be interesting to notice that the ratio between the thermal entropy ($\sim T^{d_e/z}$) and the temperature correction to entanglement entropy ($\sim T^{\frac{d_e+z}{z}}$) has a simple linear temperature relation as $\sim T$ independent of the HV parameters.

On the other hand, unlike the temperature correction, the cutoff effect depends on the parameter range. In particular, it is instructive to consider the $r_c/\ell\rightarrow0$ limit for the cutoff correction as
\begin{align}\label{cufmmew}
\begin{split}
(0 < d_e < 1):&\quad \frac{2^{d_e-1}\Gamma_1^{d_e}}{1-d_e} - \frac{1}{1-d_e} \left(\frac{r_c}{\ell}\right)^{1-d_e} \,, \\  (d_e=1):&\quad -\log{\frac{r_c}{\ell}} \,, \\ 
(d_e > 1):& \quad \frac{1}{d_e-1} \frac{1}{\left(\frac{r_c}{\ell}\right)^{d_e-1}} \,, 
\end{split}
\end{align}
In this limit, the $(0<d_e<1)$ case has a finite $S(A)$, while the other two cases $(d_e\geq1)$ are divergent. In particular, such divergence structure is also dependent of the parameter ranges: it is logarithmic for $(d_e=1)$ and has a power-law behavior for $(d_e>1)$.

Based on Eq.\eqref{cufmmew}, we also find the role of the finite cut off $r_c/\ell$: the cutoff \textit{suppresses} the entanglement entropy $S(A)$. In other words, the cutoff has  opposite corrections compared to the temperature.

%

\subsubsection{Large cutoff analysis}\label{sec3.3}

Next, following a similar procedure given in the small cutoff analysis above, we also perform the large cutoff analysis ($\tilde{r}_c \sim 1 \ll \tilde{r}_h$).
Similar to Eq.\eqref{SMCREVI}, one can also check that our large cutoff condition ($\tilde{r}_c \sim 1 \ll \tilde{r}_h$) is equivalent to  
\begin{align}\label{LCL}
   r_c \sim r_t \ll r_h    \quad\rightarrow\quad  \ell \ll r_c \ll r_h  \quad\rightarrow\quad   1 \ll r_c/\ell \ll r_h/\ell  \,.
\end{align}
for instance, from Eq.\eqref{rte2} one could find that $r_c\sim r_t$ implies $\ell \ll r_c$.

In this large cutoff limit, $S(A)$ in Eq.\eqref{HEEFOR} produces
\begin{align}\label{Sle2}
\begin{split}
    \frac{S(A)}{\ell^{1-d_e}} & = \frac{\ell^{d_e}}{r_c^{d_e}}  \Bigg[  \frac{1}{2} - \frac{d_e^2}{48}\left\{1-\left(\frac{r_c}{r_h}\right)^{d_e+z} \right\}\left( \frac{\ell}{r_c} \right)^2   +   \frac{d_e^3}{3840} \Bigg\{ (8+d_e) - 2(8-d_e-2z) \left(\frac{r_c}{r_h}\right)^{d_e+z}  \\
    & \qquad\quad + (8 - 3d_e - 4z) \left(\frac{r_c}{r_h}\right)^{2d_e+2z}  \Bigg\} \left( \frac{\ell}{r_c} \right)^4 \Bigg] \,,
\end{split}
\end{align}
where we omit higher terms $\mathcal{O}(\ell/r_c)^6$. Note that according to Eq.\eqref{LCL}, $r_c/r_h$ corresponds to the small temperature correction. One interesting observation is that unlike the small cutoff case Eq.\eqref{Sle1de}, now $S(A)$ in the large cutoff Eq.\eqref{Sle2} is independent of the parameter regime Eq.\eqref{RANGECON}.
 
Also, similar to the small cutoff, Eq.\eqref{rte1}, in order to express $S(A)$ at fixed $\ell$, we used
\begin{align}\label{rte2}
\begin{split}
    r_t =&\, r_c\Bigg[1 + \frac{d_e}{8} \left\{1 - \left(\frac{r_c}{r_h}\right)^{d_e+z} \right\} \left(\frac{\ell}{r_c}\right)^2  -\frac{d_e^2}{384} \Bigg\{ (5-2d_e) -2(5 - 4 d_e - 2 z) \left(\frac{r_c}{r_h}\right)^{d_e+z} \\
    & \qquad  + (5 - 6 d_e - 4 z) \left(\frac{r_c}{r_h}\right)^{2d_e+2z} \Bigg\} \left(\frac{\ell}{r_c}\right)^4  \Bigg]
\end{split}
\end{align}
where we have omitted higher order terms $\mathcal{O}(\ell/r_c)^6$.

{We find an interesting feature for the large cutoff case, which is distinct from the small cutoff case: the temperature correction is irrelevant to $S(A)$ when the cutoff is large. While the role of the cutoff is similar: the cutoff suppresses $S(A)$. 

For instance, from Eq.\eqref{Sle2}, one can notice that the $r_h/\ell$ term is always in sub-leading terms of $r_c/\ell$. Alternatively, this implies that one cannot find the temperature correction in the leading term of $r_c/\ell$.

In other words, for the case of the large cutoff, it is tantamount to saying that 
\begin{itemize}
    \item{The temperature is \textit{irrelevant} to $S(A)$;}
    \item{The cutoff \textit{suppresses} $S(A)$.}
\end{itemize}

\paragraph{Further comments on the temperature dependence:}
As could be seen in Fig. \ref{RTCT}, the horizon has its allowed region with the bulk parameters ($r_c, r_t$) as $r_c<r_t<r_h$. This implies that the value of $r_h$ may have a lower bound as $r_c \sim r_t \sim r_h$. In this case, one can imagine that the Ryu-Takayanagi surface  (red surface in Fig. \ref{RTCT}) may be pressed down by the Horizon so that its area would coincide with the subsystem size $\ell$ (black line in Fig. \ref{RTCT}).

Then, $S(A)$ subsequently would 
\begin{align}\label{largeT}
\lim_{r_h \,\to\, r_c} S(A) = \frac{\ell}{2r_c^{d_e}}, 
\end{align}
where its induced metric is 
\begin{align}\label{grh}
d s^{2}=\frac{d x_{1}^{2}+d x_{2}^{2}+...+d x_{d}^{2}}{r_c^{2d_e/d}} \,,
\end{align}
obtained by Eq.\eqref{HVMETRIC2} at fixed time and radius at $r=r_c$.
Thus, the temperature dependence of $S(A)$ is supposed to be placed in between Eq.\eqref{Sle1de}/Eq.\eqref{Sle2} ($r_h \rightarrow \infty$) and Eq.\eqref{largeT} ($r_h \rightarrow r_c$).

\paragraph{Numerical results vs. Analytic results:}
Based on all the analytic analysis including the small/large cutoff case, we also present some numerical plots to check the validity of our analytic results above.

In Fig. \ref{FIG11}, we make the representative plots of $S(A)$ for the \textit{cutoff} dependence.
\begin{figure*}[]
\centering
     \subfigure[($0<d_e<1$) case:\,\, $d_e=1/2$]
     {\includegraphics[width=4.8cm]{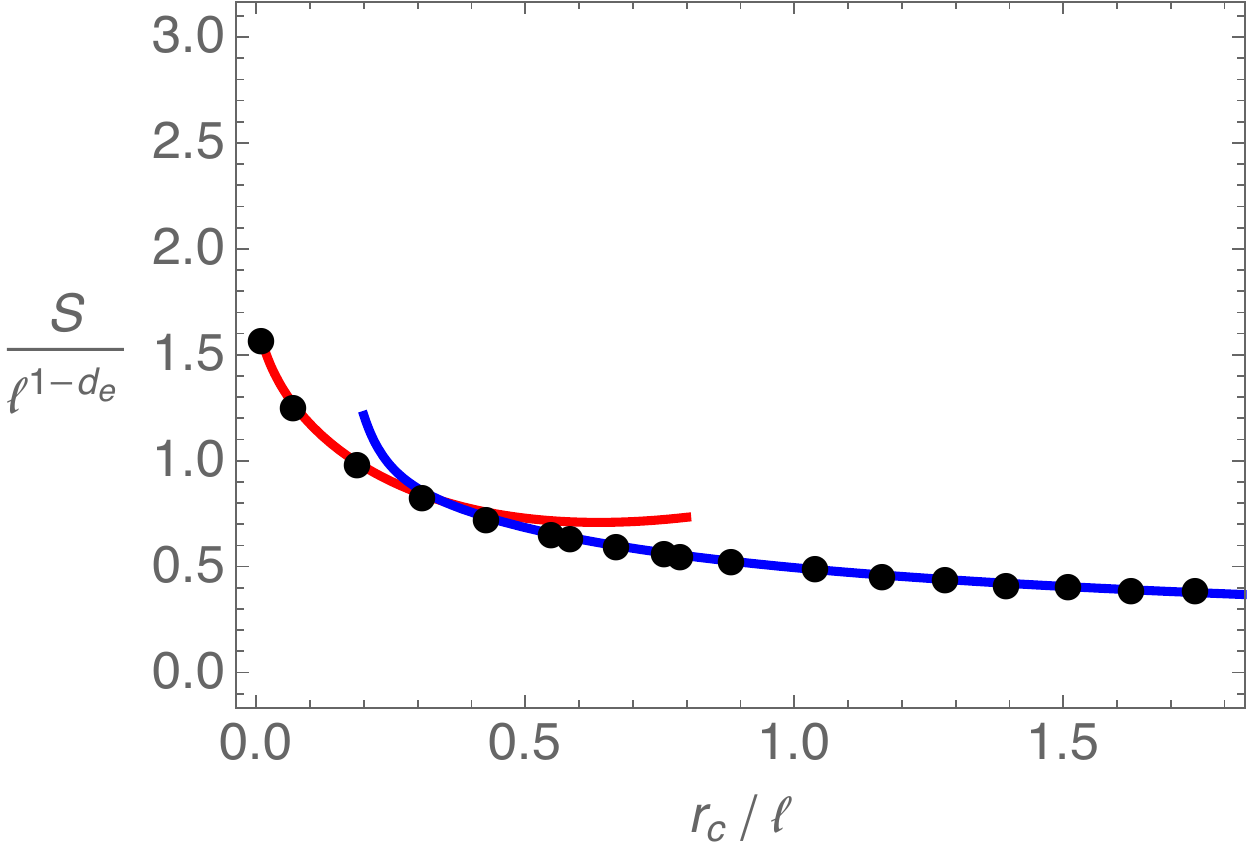} \label{}}
     \subfigure[($d_e=1$) case:\,\, $d_e=1$]
     {\includegraphics[width=4.8cm]{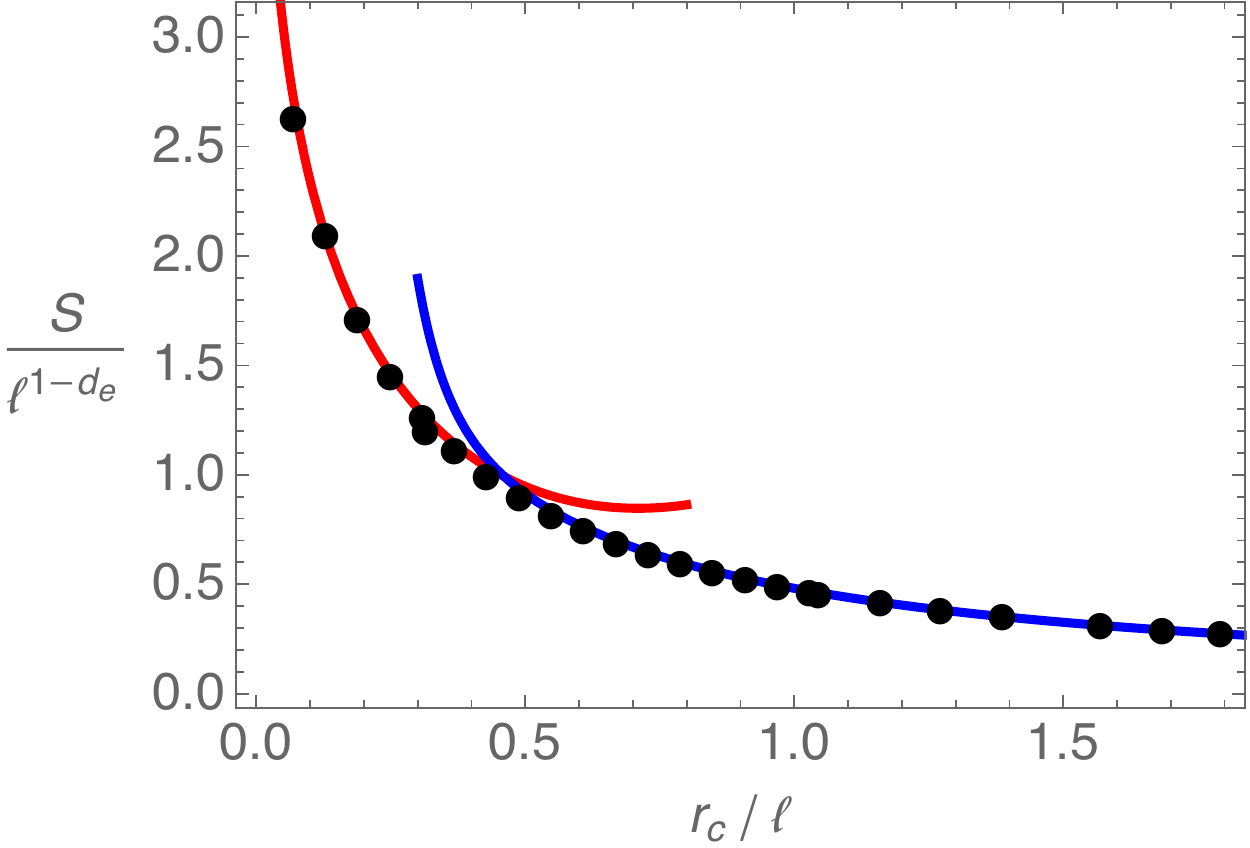} \label{}}
     \subfigure[($d_e>1$) case:\,\, $d_e=2$]
     {\includegraphics[width=4.8cm]{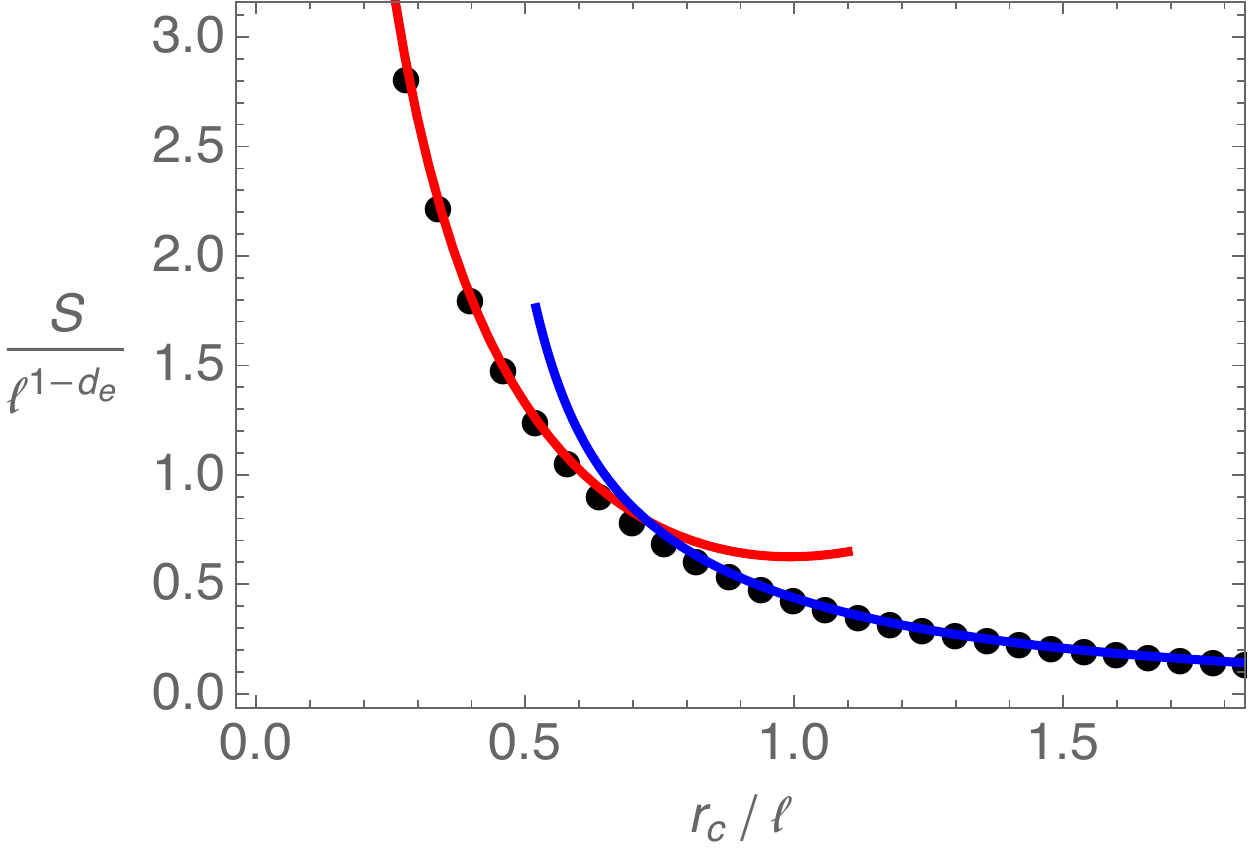} \label{}}
 \caption{$S/\ell^{1-d_e}$ vs. $r_c/\ell$ at $z=2$ and $r_h/\ell=5$. Dots are numerical results and solid lines are analytic results: Eq.\eqref{Sle1de} (red; small cutoff), Eq.\eqref{Sle2} (blue; large cutoff).}\label{FIG11}
\end{figure*}
One could see that not only the cutoff always suppresses $S(A)$ for all parameter ranges as expected, but also our analytic results are in good agreement with the numerical results: Eq.\eqref{Sle1de} for the small cutoff and Eq.\eqref{Sle2} for the large cutoff.

We also display the \textit{temperature} dependence in Fig. \ref{FIG12}.
\begin{figure*}[]
\centering
     \subfigure[$r_c/\ell=0.1$ (small cutoff)]
     {\includegraphics[width=5.8cm]{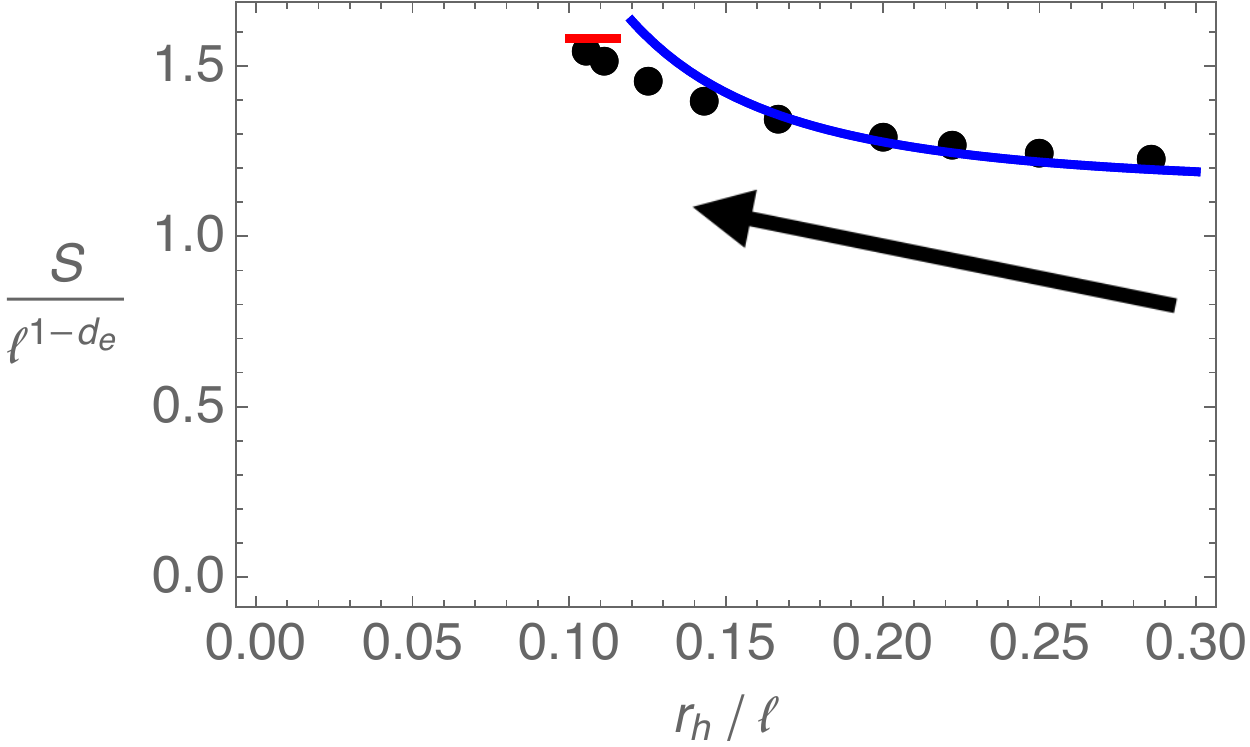} \label{}}
     \subfigure[$r_c/\ell=2$ (large cutoff)]
     {\includegraphics[width=5.8cm]{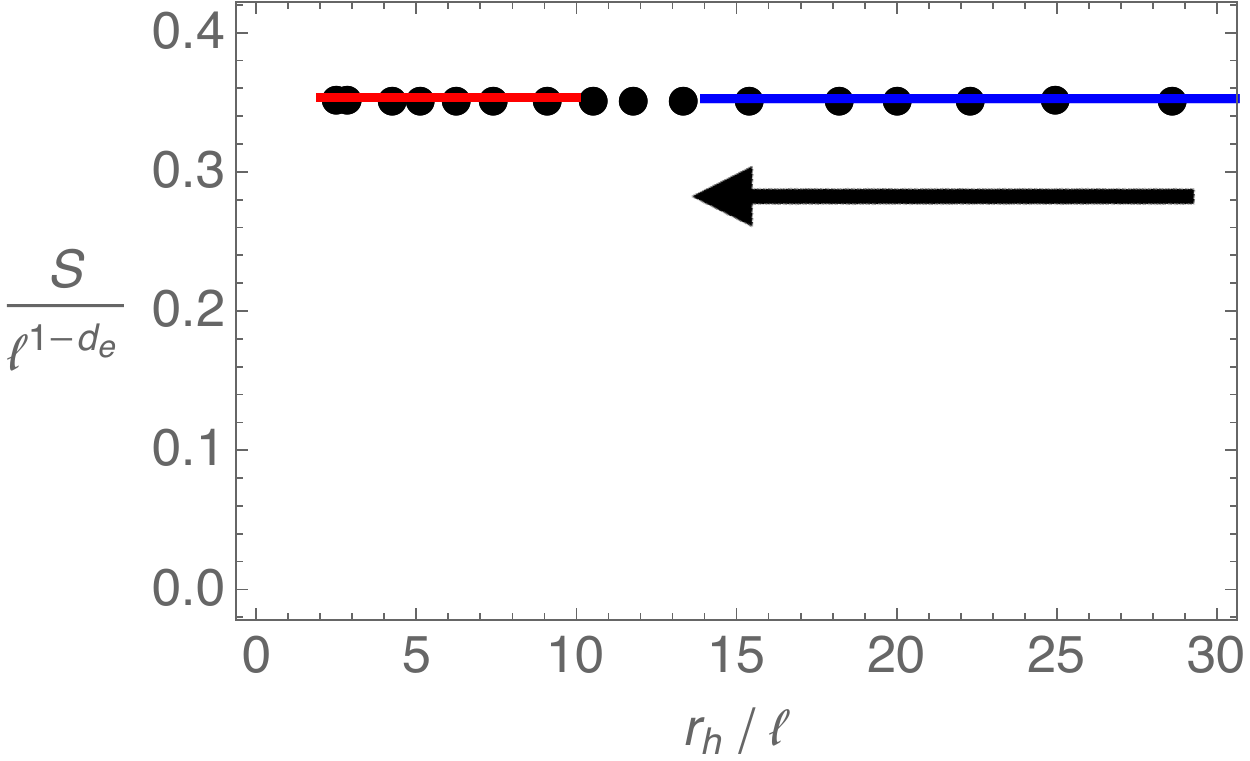} \label{}}
 \caption{$S/\ell^{1-d_e}$ vs. $r_h/\ell$ at $(d_e, z)=(1/2, 2)$. Dots are numerical results interpolating the analytic results between Eq.\eqref{largeT} (red; $r_h \rightarrow r_c$) and Eq.\eqref{Sle1de} (blue; $r_h \rightarrow \infty$; left panel) or Eq.\eqref{Sle2} (blue; $r_h \rightarrow \infty$; right panel). In all figures, the arrow indicates the direction of increasing temperature.}\label{FIG12}
\end{figure*}
Recall that the temperature correction is universal in the parameter range: Eq.\eqref{UTCF} for the small cutoff, Eq.\eqref{Sle2} for the large cutoff.
One could check that the temperature enhances $S(A)$ for the small cutoff case (the left panel of Fig. \ref{FIG12}), while it is irrelevant for $S(A)$ in the large cutoff case (the right panel of Fig. \ref{FIG12}). 
Also notice that $S(A)$ approaches Eq.\eqref{largeT} when $r_h \rightarrow r_c$.

}

\section{Some applications}
\label{sec4}

\subsection{First law of entanglement entropy}\label{sec3.4}

One of the well known applications for the study of the holographic entanglement entropy at \textit{finite} temperature is the first law of entanglement entropy (FLEE)~\cite{Bhattacharya:2012mi}:
\begin{align}\label{FLEE}
    \delta S(A)= \frac{\delta E(A)}{T_{\text{ent}}},
\end{align}
where $T_{\text{ent}}$ is the entanglement temperature, $\delta S(A)$ the increased amount of the entanglement entropy in the thermal state ($T\neq0$) compared to the ground state of the CFT ($T=0$):
\begin{align}\label{FLEET}
    \delta S(A) := S(A)_{T}-S(A)_{T=0}.
\end{align}
$\delta E$ is the corresponding increased energy in the subsystem given by $\int d^{d}x \, T_{xx} = \ell L \, T_{xx} \sim \ell$. 

Note that FLEE is only valid for the small subsystem $\ell$ as  
\begin{align}\label{}
    \delta S(A) \sim T_{\text{ent}}^{-1} \,\, \ell \,,
\end{align}
where $T_{\text{ent}}$ may not depend on the details of the excited states. For instance, for the pure AdS ($d_e=d, z=1$), it is known to be $T_{\text{ent}}^{-1} \sim \ell$ so that $\delta S(A) \sim \ell^2$.
Furthermore, when the cutoff is zero, $r_c\rightarrow0$, it was also shown \cite{Bhattacharya:2012mi} that 
\begin{align}\label{zcrpr}
   T_{\text{ent}} \sim \ell^{-z} \quad\rightarrow\quad \delta S(A) \sim \ell ^{1+z} \,,
\end{align}
for the HV geometry.

Here, we discuss FLEE at finite cutoff for the HV geometry. Unlike the zero cutoff, for the finite cutoff case, we may have two cases describing the ``small" subsystem ($\ell \ll r_h$)
\begin{align}\label{}
   r_c \ll \ell \ll r_h\,, \qquad \ell \ll r_c \ll r_h\,,
\end{align}
where the former case is what we called the small cutoff in section \ref{sec3.2} and the latter the large cutoff in section \ref{sec3.3}.
Notice that the zero cutoff case~\cite{Bhattacharya:2012mi} is included in the former one.

Let us first discuss the \textit{small} cutoff case. In this case, one could find that $\delta S(A)$ is the parameter-range-independent temperature correction, Eq.\eqref{UTCF}, 
\begin{align}\label{deltaSe1}
\delta S(A)= \frac{d_e \Gamma_2}{2^{2+z}(1+z)(1+d_e+z)\Gamma_1^{1+z}}\left(\frac{\ell^{1+z}}{r_h^{d_e+z}}\right) \sim \ell^{1+z}\,,
\end{align}
which is consistent with Eq.\eqref{zcrpr}. 

On the other hand, for the \textit{large} cutoff case, based on our analytic result Eq.\eqref{Sle2}, we find the leading contribution of temperature as
\begin{align}\label{deltaSe2}
\delta S(A)=\frac{d_e^2}{48}\frac{r_c^{z-2}}{r_h^{d_e+z}} \, \ell^3 \sim \ell^3 \,,
\end{align}
which is an interesting observation since the power of $\ell$ turns out to be a universal value independent of $d_e$ or $z$ unlike the small cutoff case Eq.\eqref{deltaSe1}.

\begin{figure*}[]
\centering
     \subfigure[$z$-effect: $d_e=1, z \in \text{[1, 3]} $ (red-blue)]
     {\includegraphics[width=5.8cm]{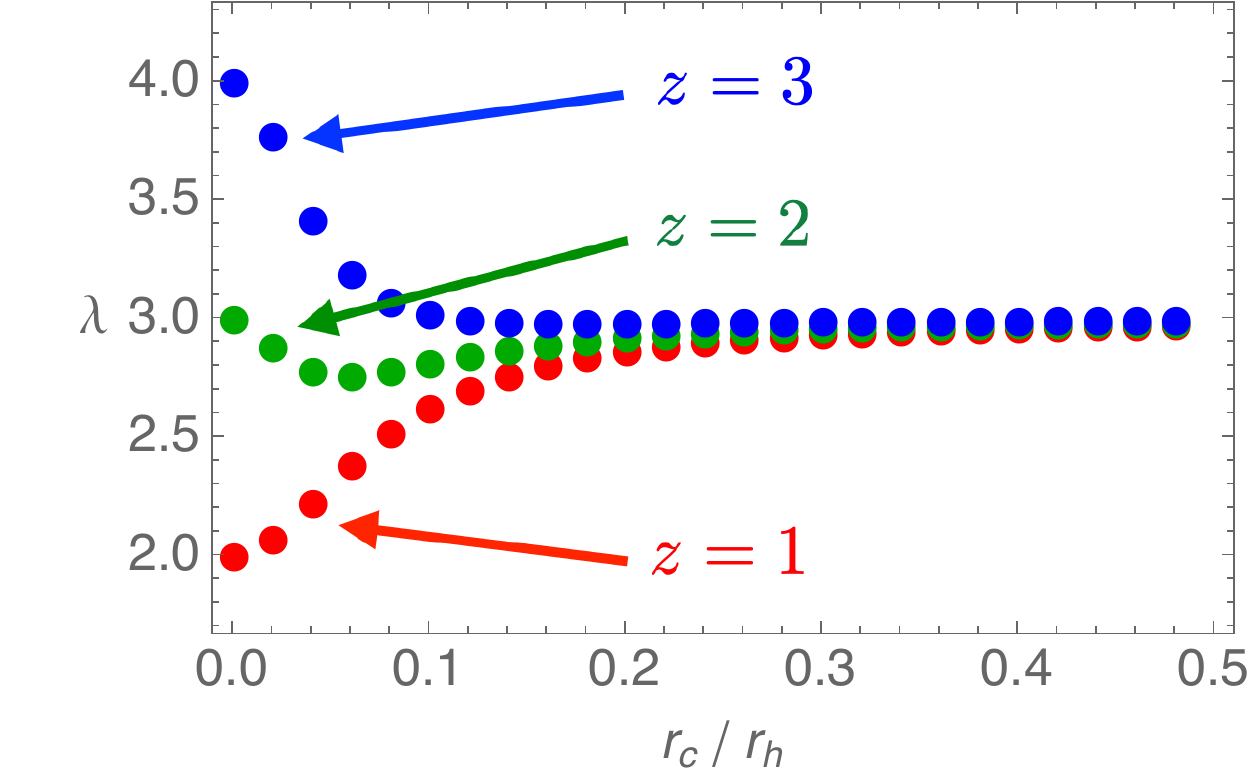} \label{}}
     \subfigure[$d_e$-effct: $d_e \in \text{[1, 3]}$ (red-blue), $z=3$]
     {\includegraphics[width=5.8cm]{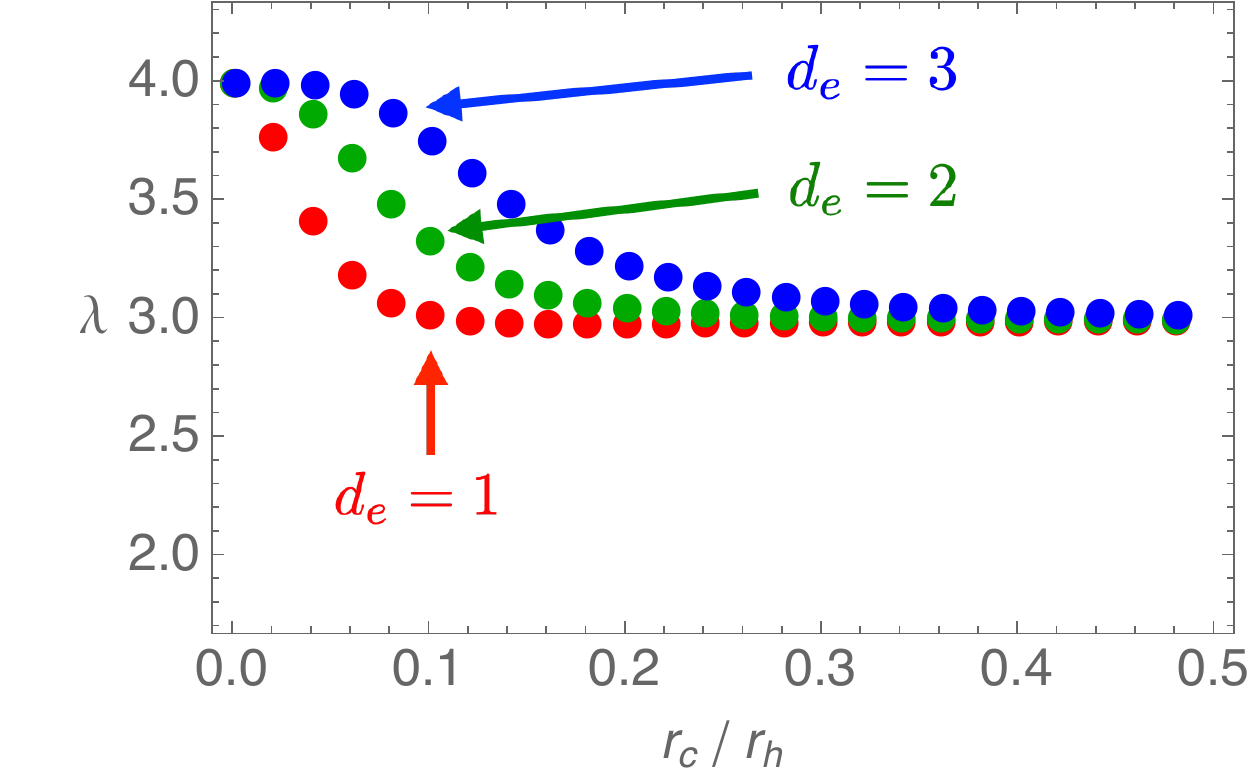} \label{}}
 \caption{The power $\lambda$ vs. $r_c/r_h$. The numerical result interpolating two (analytic) limiting cases: $\lambda = 1+z$ in Eq.\eqref{deltaSe1} at small cutoff and $\lambda = 3$ in Eq.\eqref{deltaSe2} at large cutoff.}\label{figFLEE}
\end{figure*}

One could also study FLEE beyond the small/large cutoff case, for instance, the intermediate case as 
\begin{align}\label{}
   r_c \sim \ell \ll r_h \,.
\end{align}
In particular, by fitting the numerical result with the power law behavior
\begin{align}\label{}
   \delta S(A) \,\sim\, \ell^\lambda \,,
\end{align}
one can find the behavior of $\lambda$.
We display its plot in Fig. \ref{figFLEE}
where the result interpolates between two (analytic) limiting cases: Eq.\eqref{deltaSe1} at small cutoff and Eq.\eqref{deltaSe2} at large cutoff.

%
\subsection{Analytic gravity dual of $S(A)$ for $T\bar{T}$ like deformed HV QFTs: $d_e=z$}\label{sec3.5}

We close this section by providing the analytic gravity result of $S(A)$, which is valid for arbitrary value of bulk parameters ($r_c, r_h, r_t$), i.e., beyond the small/large cutoff regime.

Choosing the special combination between the HV parameters as $d_e=z$, we find the entanglement entropy Eq.\eqref{HEEFOR} analytically, $S^{\text{Gravity}}_{d_e=z} := S(A)\big|_{d_e=z}$, as
\begin{align}\label{GDHV}
\begin{split}
    S^{\text{Gravity}}_{d_e=z} =  &\,\, \frac{r_t^{1-d_e}}{1-d_e}\Bigg[ F_1\left( {\frac{1-d_e}{2d_e}; \frac{1}{2},\,\frac{1}{2};\,\frac{1+d_e}{2d_e};\,\frac{r_t^{2d_e}}{r_h^{2d_e}},\,1} \right) \\
    & \qquad -\left(\frac{r_c}{r_t}\right)^{1-d_e} F_1\left(  { \frac{1-d_e}{2d_e};\,\frac{1}{2},\,\frac{1}{2};\,\frac{1+d_e}{2d_e};\,\frac{r_c^{2d_e}}{r_h^{2d_e}},\,\frac{r_c^{2d_e}}{r_t^{2d_e}} }\right) \Bigg]
\end{split}
\end{align}
as well as the subsystem size Eq.\eqref{lengthfor} as
\begin{align}\label{lengthGDHV}
\begin{split}
   \ell &= \,\, \frac{2 r_t}{1+d_e}\Bigg[ F_1\left( {\frac{1+d_e}{2d_e};\,\frac{1}{2},\,\frac{1}{2};\,\frac{1}{2}\left(3+\frac{1}{d_e}\right);\,\frac{r_t^{2d_e}}{r_h^{2d_e}},1} \right)
    \\
    & \qquad\quad -\frac{r_c^{1+d_e}}{r_t^{1+d_e}} F_1\Bigg( {\frac{1+d_e}{2d_e};\,\frac{1}{2},\,\frac{1}{2};\,\frac{1}{2}\left(3+\frac{1}{d_e}\right); \,\frac{r_c^{2d_e}}{r_h^{2d_e}},\,\frac{r_c^{2d_e}}{r_t^{2d_e}}} \Bigg) \Bigg]
\end{split}
\end{align}
where $F_1$ is the Appell hypergeometric function.

We also find that the allowed parameter range with $d_e=z$ becomes 
\begin{align}\label{ads3prd}
\begin{split}
d_e \,=\,z \,\geq\, \frac{2d}{1+d} \,,
\end{split}
\end{align}
which belongs to the third case in Eq.\eqref{RANGECON}.
Notice that the known gravity dual of $S(A)$ for the $T\bar{T}$ deformed CFTs, ($d_e=z=1$), is only viable at $d=1$, i.e., $AdS_3$ black holes~\cite{Donnelly:2018bef,Chen:2018eqk,Park:2018snf,Banerjee:2019ewu,Murdia:2019fax,Jeong:2019ylz,Grieninger:2019zts,Donnelly:2019pie}, which is consistent with Eq.\eqref{ads3prd}.

Furthermore, we make two comments for the gravity result Eq.\eqref{GDHV} as
\begin{itemize}
    \item{The condition $d_e=z$ can only be implemented at finite $T$ (or $f(r)\neq1$) in that the dynamical critical exponent $z$ only appears in $S(A)$ when $f(r)\neq1$, i.e., Eq.\eqref{GDHV} can be regarded as an another application of the finite temperature analysis.}
    \item{If one can obtain the entanglement entropy of $T\bar{T}$ like deformed QFTs with ($d_e=z$), our result Eq.\eqref{GDHV} could be used as the prediction for its gravity dual, i.e., Eq.\eqref{GDHV} can be used for checking holographic duality in the future.\footnote{Note that it would be important to have the full analytic gravity result for checking the holographic duality in that, for instance within the AdS$_3$ case, the duality has been reported at small cutoff and high temperature, which is beyond our scope (small cutoff, low temperature).}}
\end{itemize}
For instance, for $T\bar{T}$ like deformed QFTs with $d_e=z$, the field theory parameters would be ($\xi, T, \ell$); recall that $\xi$ is a deformation parameter in Eq.\eqref{OPX}.
Thus, such a field theory result of the entanglement entropy, $S^{\text{QFT}}_{d_e=z} \left(\xi\,,T\,,\ell\right)$, could be compared with our gravity result Eq.\eqref{GDHV} as
\begin{align}
\begin{split}
S^{\text{QFT}}_{d_e=z} \left(\xi\,,T\,,\ell\right) \quad\Longleftrightarrow\quad S^{\text{Gravity}}_{d_e=z}\left(r_c\,,r_h\,,r_t\right) \,,
\end{split}
\end{align}
when we use $\xi(r_c)$ Eq.\eqref{OPX} and $T(r_h)$ from Eq.\eqref{HAWKT} together with $\ell(r_c, r_h, r_t)$ Eq.\eqref{lengthGDHV}.

\section{Conclusion}\label{sec5}
We have studied the holographic entanglement entropy of the Hyperscaling Violating (HV) geometries at finite temperature and finite cutoff for which such a gravity setup is conjectured to be dual to the $T\bar{T}$ like deformed HV QFTs~\cite{Alishahiha:2019lng}.

This paper is along the line of the recent development to reveal the quantum information features of the $T\bar{T}$ like deformed HV QFTs from the perspective of gravity: holographic complexity~\cite{Alishahiha:2019lng}, holographic entanglement entropy, mutual information, and entanglement wedge cross section~\cite{Khoeini-Moghaddam:2020ymm,Paul:2020gou}. 

In particular, considering the black hole horizon geometry, we have analytically studied the finite temperature/cutoff effect on the entanglement entropy with full HV parameters ($d_e\,, z$) within its complete parameter range Eq.\eqref{RANGECON}. Also we have checked that our analytic treatment is in good agreement with numerical results.

Recall that our work not only extends the previous zero temperature result~\cite{Khoeini-Moghaddam:2020ymm,Paul:2020gou} to the finite temperature case in order to study the $z$ contribution, but also fills the gap by considering all possible parameter range Eq.\eqref{RANGECON}.

We summarize the main results of the temperature/cutoff effects as:
\begin{itemize}
    \item{ The temperature \textit{enhances} $S(A)$ Eq.\eqref{UTCF} in the {small} cutoff limit, while it is \textit{irrelevant} in the {large} cutoff limit  Eq.\eqref{Sle2};}
    \item{ The cutoff \textit{suppresses} $S(A)$: Eq.\eqref{cufmmew}/Eq.\eqref{Sle2} in the small/large cutoff limit;}
\end{itemize}
and we also find the parameter range dependence as
\begin{itemize}
    \item{ The temperature correction is independent of the parameter range: Eq.\eqref{UTCF}, Eq.\eqref{Sle2};}
    \item{ The cutoff correction could depend on the parameter range: Eq.\eqref{cufmmew}.}
\end{itemize}

Furthermore, as an application of our finite temperature analysis, we also studied the first law of entanglement entropy (FLEE). We find that the FLEE of the HV geometries in the presence of the cutoff can be expressed as 
\begin{align}\label{}
\delta S(A) \sim \ell^{1+z} \, (\text{small cutoff}) \,,\quad \delta S(A)\sim \ell^3 \, (\text{large cutoff}) \,,
\end{align}
where they are functions depending only on the dynamical critical exponent $z$, similar to the case of the universal bound of diffusion constants~\cite{Blake:2016sud,Blake:2016jnn,Blake:2017qgd,Ahn:2017kvc}. 

We also provide, as another application at finite temperature, the analytic gravity dual for the  entanglement entropy of $T\bar{T}$ like deformed QFTs with ($d_e=z$): Eq.\eqref{GDHV}.

It would be interesting to compute the entanglement entropy for the deformed HV QFTs and check if one could find similar temperature/cutoff features that we found and also explicitly compare it with our gravitational result. As suggested in \cite{Khoeini-Moghaddam:2020ymm,Paul:2020gou}, in general the usual CFT techniques may not be simply applicable for the HV QFTs in that its symmetry group is smaller than the conformal group. However, when the boundary manifold on which the QFT lives is maximally symmetric such as $S^{d+1}$, one may still have some possibility for a reliable QFT computation. We hope to address this direction in the near future.

\acknowledgments

We would like to thank {Mitsuhiro Nishida, Shao-Hua Xue}  for valuable discussions and correspondence.
This work was supported by the National Key R$\&$D Program of China (Grant No. 2018FYA0305800), Project 12035016 supported by National Natural Science Foundation of China, the Strategic Priority Research Program of Chinese Academy of Sciences, Grant No. XDB28000000.
H.-S Jeong acknowledges the hospitality at Gwangju Institute of Science and Technology (GIST) where part of this work was done.
W. Pan and Y. Wang contributed equally to this paper and should be considered co-first authors.

\appendix
%


\bibliographystyle{JHEP}

\providecommand{\href}[2]{#2}\begingroup\raggedright\endgroup

\end{document}